\documentclass[runningheads]{llncs}
\usepackage[T1]{fontenc}
\usepackage{graphicx}
\usepackage{booktabs,multirow}
\usepackage{amsmath}
\usepackage{xcolor} 
\usepackage[misc]{ifsym}
\newcommand{\corr}{(\Letter)}
\usepackage{mwe}

\begin{document}

\title{Hierarchical Interaction Summarization and Contrastive Prompting for Explainable Recommendations}
\titlerunning{PGHIS-CPEG}


\author{Yibin Liu$^{1*}$ \and
Ang Li$^{1*}$ \and
Shijian Li\corr$^{1,2}$
}

\institute{
$^1$College of Computer Science and Technology, Zhejiang University, Hangzhou 310027, China\\
$^2$ the State Key Lab of Brain-Machine Intelligence\\
\email{\{yibinliu, leeyon, shijianli\}@zju.edu.cn}
}

\maketitle              
\def\thefootnote{*}\footnotetext{Equal Contribution}

\begin{abstract}
Explainable recommendations, which use the information of user and item with interaction to generate a explanation for why the user would interact with the item, are crucial for improving user trust and decision transparency to the recommender system. Existing methods primarily rely on encoding features of users and items to embeddings, which often leads to information loss due to dimensionality reduction, sparse interactions, and so on. With the advancements of large language models (LLMs) in language comprehension, some methods use embeddings as LLM inputs for explanation generation. However, since embeddings lack inherent semantics, LLMs must adjust or extend their parameters to interpret them, a process that inevitably incurs information loss. To address this issue, we propose a novel approach combining profile generation via hierarchical interaction summarization (PGHIS), which leverages a pretrained LLM to hierarchically summarize user-item interactions, generating structured textual profiles as explicit representations of user and item characteristics. Additionally, we propose contrastive prompting for explanation generation (CPEG) which employs contrastive learning to guide another reasoning language models in producing high-quality ground truth recommendation explanations. Finally, we use the textual profiles of user and item as input and high-quality explanation as output to fine-tune a LLM for generating explanations. Experimental results on multiple datasets demonstrate that our approach outperforms existing state-of-the-art methods, achieving a great improvement on metrics about explainability (e.g., 5\% on GPTScore) and text quality(e.g., 20.6\% and 19.6\% on variants of BLEU and ROUGE). Furthermore, our generated ground truth explanations achieve a significantly higher win rate compared to user-written reviews and those produced by other methods, demonstrating the effectiveness of CPEG in generating high-quality ground truths.

\end{abstract}

\section{Introduction}
Recommender systems play a crucial role in delivering personalized content across domains like e-commerce, streaming services, and social media\cite{DBLP:journals/jbd/RoyD22}. They enhance user experience and engagement by analyzing past interactions and preferences. Although contemporary recommender systems achieve high accuracy in their suggestions, the generation of explanations remains underexplored. Explainable recommendations, which use the information of the user and the item with interaction to generate an explanation for why the user would interact with the item, are essential as they provide transparency into the recommendation process, enhancing user trust and facilitating informed decision-making. By elucidating the factors driving user-item interactions, these explanations bridge the gap between complex algorithms and user comprehension, thereby fostering more trustworthy and user-centric recommendation platforms.

Early research\cite{DBLP:conf/eacl/ZhouLWDHX17,DBLP:conf/sigir/LiWRBL17,DBLP:conf/acl/LiZC20} predominantly leveraged conventional natural language generation methods, including LSTM\cite{DBLP:conf/nips/ShiCWYWW15}, GRU\cite{DBLP:journals/corr/ChungGCB14}, and transformers\cite{DBLP:conf/nips/VaswaniSPUJGKP17}, to learn ID embeddings for users and items and subsequently generate corresponding explanations. With the advancements of LLMs in language comprehension, recent studies\cite{DBLP:journals/tois/LiZC23,DBLP:conf/emnlp/MaR024} have utilized these models to understand and learn embeddings of users and items to generate explanations. 

\begin{figure}[t]
\includegraphics[draft=false,width=\textwidth]{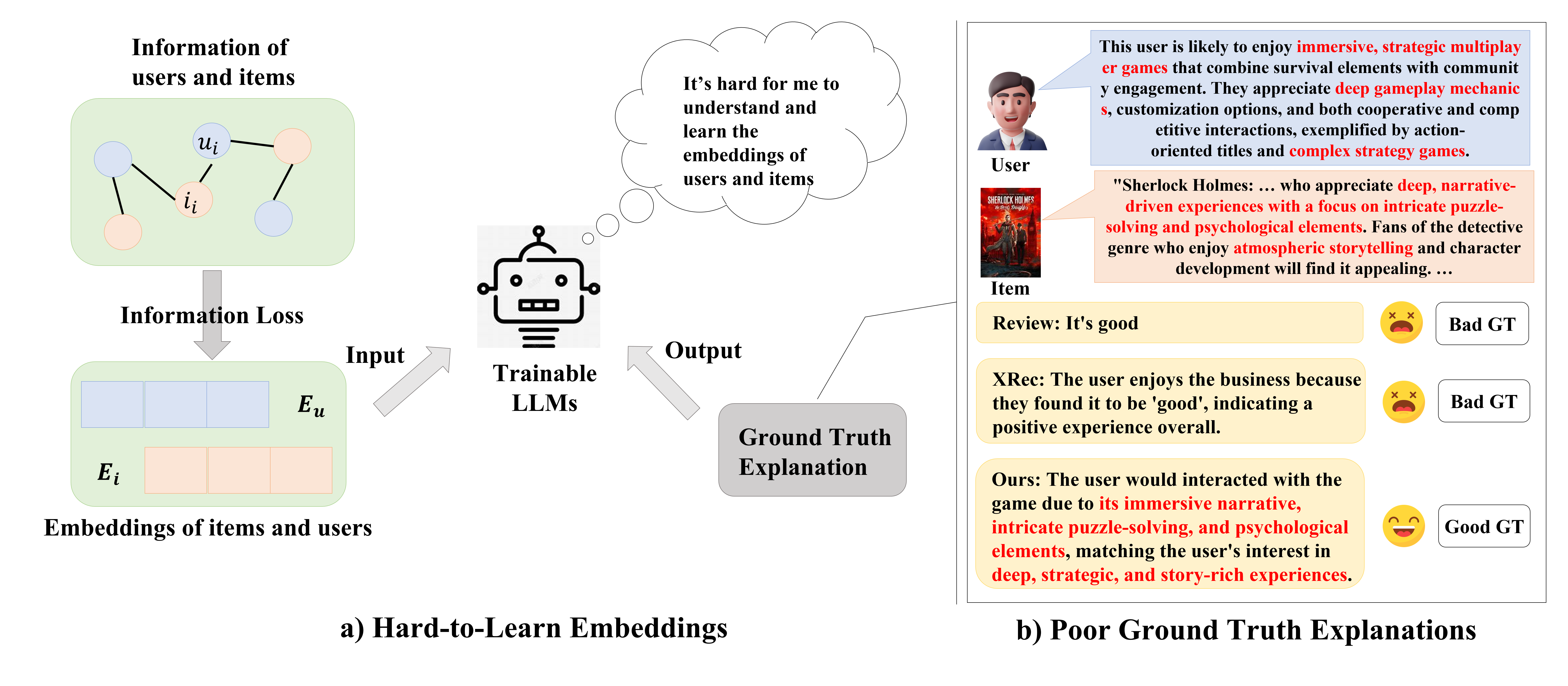}
\caption{Problems in Recommendation Explanation Generation. a) Hard-to-Learn Embeddings, learned user or item embeddings have information loss and are hard for LLMs to interpret; b) Poor ground truth Explanations, as user reviews are often simplistic, leading to low-quality explanations.} \label{motivation}
\end{figure}

Although existing methods have made significant breakthroughs in generating recommendation explanations, as shown in Figure \ref{motivation}, they still face two main problems. \textbf{a) Hard-to-Learn Embeddings}, whether learning user and item embeddings from scratch\cite{DBLP:conf/eacl/ZhouLWDHX17,DBLP:conf/sigir/LiWRBL17,DBLP:conf/acl/LiZC20,DBLP:journals/tois/LiZC23} or employing models to transform existing embeddings\cite{DBLP:conf/emnlp/MaR024} within the recommender system, these approaches often incur information loss\cite{DBLP:journals/corr/abs-2109-09587} during the embedding training or transformation processes, exacerbated by dimensionality reduction, sparse interactions, and so on. In LLM-based methods, when an LLM takes embeddings as input, adjusting the parameters of LLMs or incorporating additional parameters is required for interpretation, also leading to potential information loss. \textbf{b) Poor Ground Truth Explanation}, most current approaches use user reviews as ground truth for recommendation explanations. However, for users who do not frequently provide detailed reviews, their reviews are often simplistic, such as ``it's good'', which lack sufficient explanatory information, and thus becoming low-quality explanations in the dataset. Therefore, we face two main challenges when designing our method: (1) How can we represent information of user and item as LLM inputs in a textual format that preserves both interaction and semantic information without relying on embeddings? (2) How can we generate high-quality ground truth explanations for training, especially when user reviews are lack detailed information?


To tackle these issues, we propose \textbf{Profile Generation via Hierarchical Interaction Summarization (PGHIS)}, leveraging a pretrained LLM to hierarchically summarize interaction data, reducing information loss. Specifically, we enable the LLM to simulate graph neural networks from a textual perspective, summarizing key interactions between items and users. Through multi-layer iterations, it extracts shared attributes from interacted item-user pairs, ultimately generating structured textual profiles. Additionally, we introduce \textbf{Contrastive Prompting for ground truth Explanation Generation (CPEG)} to enhance the quality of training data. We prompt Reasoning Language Models (RLMs)\cite{besta2025reasoning} with a user and both positive and negative items to infer interactions and generate explanations. If the item predicted by the RLM is positive item, the explanation is considered high-quality. Otherwise, the erroneous explanation and item are treated as a negative example for the RLM, thereby enhancing the quality of ground-truth explanations. Finally, we construct training datasets using generated profiles as input and refined explanations as output, applying Supervised Fine-Tuning (SFT) on a pretrained LLM for our final model. Experiments on diverse datasets show high-quality explanations and significant improvements, with an average gain of 5\% on GPTScore, 20.6\% and 19.6\% on variants of BLEU and ROUGE. Our main contributions can be summarized as follows:
\begin{itemize}
    \item We identify embedding limitations and ground truth quality as key barriers in explainable recommendation, as embeddings often suffer from information loss, and lack of inherent semantics in LLM-based methods.
    \item We propose a framework that integrates hierarchical interaction summarization to create textual user-item profiles instead of embeddings and contrastive prompting to generate high-quality ground truth explanations.
    \item We evaluated our approach against state-of-the-art baselines on diverse datasets, achieving an average improvement of 5\% on GPTScore, 20.6\% and 19.6\% on variants of BLEU and ROUGE. Experimental results further validate the high quality of ground truth explanations generated by CPEG.
\end{itemize}

\section{Related Work and Preliminaries}
\subsection{Explainable Recommendation}
The primary goal of Explainable Recommendation is to generate clear textual explanations that elucidate the reasoning behind each recommendation. Given an interacted user-item pair $u$ and $i$, an explanation is generated based on their respective information $\mathcal{X}_u$ and $\mathcal{X}_i$(e.g., profiles, embeddings, or attributes), and can be described as follows:
\begin{equation}
    explanation(u, i) = Generate(\mathcal{X}_u, \mathcal{X}_i)
\end{equation}

As a natural language generation (NLG) task, existing approaches primarily leverage LSTM, GRU, and Transformer-based models\cite{DBLP:conf/eacl/ZhouLWDHX17,DBLP:conf/sigir/LiWRBL17,DBLP:conf/acl/LiZC20}. For example, Att2Seq\cite{DBLP:conf/eacl/ZhouLWDHX17} employs an attention mechanism to model the relationships between user and item attributes, guiding an LSTM to produce explanations. Similarly, NRT\cite{DBLP:conf/sigir/LiWRBL17} jointly predicts user ratings and generates concise textual tips by integrating collaborative filtering with a GRU as decoder. With the success of pretrained LLMs in NLG, some methods\cite{DBLP:journals/tois/LiZC23,DBLP:conf/emnlp/MaR024} have explored their use for this task. For example, PEPLER\cite{DBLP:journals/tois/LiZC23} initially freezes the pretrained LLM to learn user and item embeddings, and subsequently fine-tunes the LLM parameters based on the acquired embeddings. Similarly, XRec\cite{DBLP:conf/emnlp/MaR024} trains a Mixture-of-Experts
 (MOE) model\cite{DBLP:conf/kdd/HouMZLDW22} to transform the embeddings learned by the recommendater system into inputs for a pretrained LLM, thereby generating explanations. Additionally, XRec alse employs LLMs to rewrite reviews as ground truth explanation to eliminate subjectivity. However, these approaches encode user and item features into embeddings, which often suffer from information loss. This issue is exacerbated when embeddings are used as LLM inputs, as LLMs struggle to interpret floating-point embeddings, leading to further loss. In contrast, our approach shifts focus from item and user embeddings to their textual profiles.

\subsection{Graph Collaborative Filtering(GCF)}
Graph Neural Networks effectively model collaborative relationships by capturing high-order dependencies in user-item interactions. Through iterative message passing, nodes aggregate information from neighbors to generate embeddings that reflect these relationships. Given a user-item interaction graph with $L$ layers, the embedding of a user node $u$ or an item node $i$ at layer $l$ is computed as follows:
\begin{equation}
    e_u^l = AGG(e_u^{l-1}, \mathcal{N}_u)
\end{equation}
Here, $\mathcal{N}_u$ represents the set of embeddings for items that have interacted with user $u$. $AGG$ denotes an aggregation function, which can vary depending on the model. For example, NGCF\cite{NGCF19} employs a combination of a summation function and a nonlinear activation function, while LightGCN\cite{DBLP:conf/sigir/0001DWLZ020} adopts a simple summation function. In contrast, AutoCF\cite{autocf2023} introduces a masked graph autoencoder designed to aggregate global information. Similar to these methods, we propose PGHIS, which uses a pretrained LLM as the aggregation function to integrate node features, replacing embeddings with explicit textual representations of user and item characteristics. To the best of our knowledge, we are the first to apply graph neural network concepts to profile generation.

\subsection{Contrastive Learning in Prompts}
Although LLMs have achieved significant progress across various domains, they still struggle in certain scenarios. To address this, some studies have incorporated contrastive learning into prompts by providing both correct and incorrect examples\cite{li2024learning,DBLP:journals/corr/abs-2311-09277,DBLP:journals/corr/abs-2412-18925}, guiding the model to produce outputs aligned with the correct examples while avoiding errors. For instance, CCoT\cite{DBLP:journals/corr/abs-2311-09277} introduces both valid and flawed thinking steps to steer the model toward correct results. Similar to these approaches, we propose CPEG to guide LLMs in generating high-quality ground truth explanations.

\section{Method}
In this work, we propose a novel Profile Generation method via Hierarchical Interaction Summarization to generate user and item profiles by iteratively aggregating the common features of a user's or item’s interacted items or users across multiple layers, forming refined textual profiles. This eliminates the need for LLMs to interpret embedding-based interactions, as preference patterns are explicitly represented in natural language, which enhances the explainability of recommendation systems while effectively capturing user and item characteristics.

To address the limitation of user reviews in explaining why users interact with items, we propose a method called Contrastive Prompting for ground truth Explanation Generation that leverages RLMs and employs Contrastive Prompts to generate more coherent and insightful ground truth recommendation explanations. 

Finally, we use the generated user and item profiles as inputs to the model and produce recommendation explanations generated by CPEG as the ground truth output to construct training and testing datasets for our model. Then, we fine-tune a lightweight LLM using SFT with these datasets, resulting in our final recommendation explanation generation model.
\begin{figure}[t]
\centering
\includegraphics[draft=false,width=0.9\textwidth]{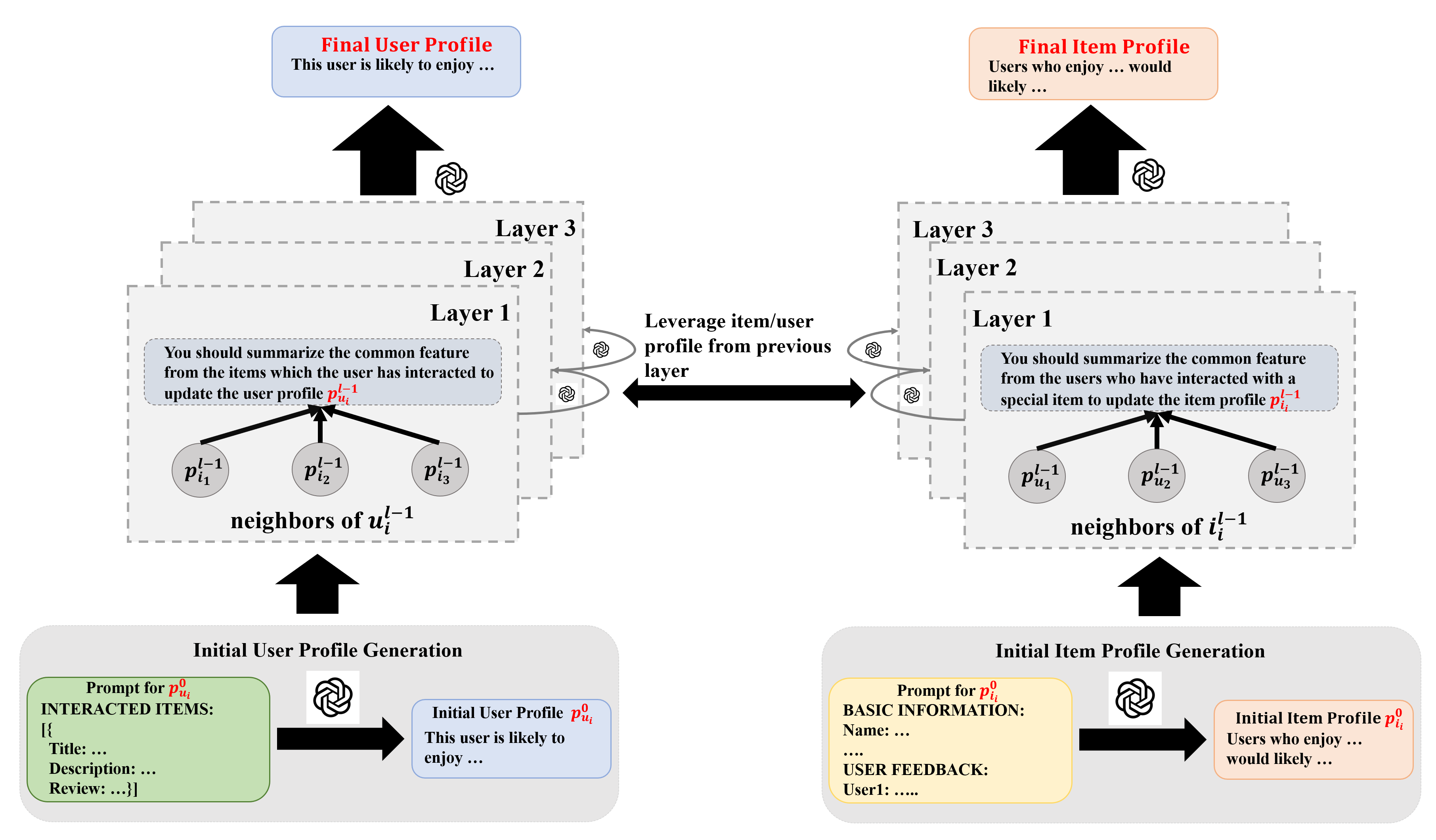}
\caption{An overall overview of the Profile Generation method via Hierarchical Interaction Summarization.} \label{prf_gen}
\end{figure}

\subsection{Profile Generation via Hierarchical Interaction Summarization}
To solve the issue of hard-to-learn embeddings, we introduce a Profile Generation method via Hierarchical Interaction Summarization, which leverages a pretrained large model as the aggregation function. This method continuously distills and summarizes the shared characteristics of items-users interacting with a given user or item, dynamically refining the original user or item profile. By doing so, it directly embeds interaction information into the profile, enhancing its expressiveness. The proposed Profile Generation method via Hierarchical Interaction Summarization is illustrated in Figure \ref{prf_gen}.

For any given number of profile updates $l$, we assume that the profile of any user or item before the update is $p_{u_i}^{l-1}$ or $p_{i_i}^{l-1}$ and the set of the profiles of items or users that have interacted with the user or item is $\mathcal{N}_{i_i}^{l-1}$ or $ \mathcal{N}_{u_i}^{l-1}$. Given an existing user or item profile, we leverage the profiles of the interacted item-user to generate an updated profile. Specifically, we use the following system prompt $\mathcal{P}_{agg}$:

\textit{"You will serve as an assistant to help me update the user/item profile ... You should summarize the common features of the interacted items/users and then update the original user/item profile."}

This prompt, along with the pretrained LLM, is used to aggregate the relevant information and derive the new user or item profile $p_{u_i}^{l}/p_{i_i}^{l}$. The update process is formulated as follows:
\begin{align}
p_{u_i}^{l} = LLM(\mathcal{P}_{agg}, p_{u_i}^{l-1}, \mathcal{N}_{i_i}^{l-1}) \\
p_{i_i}^{l} = LLM(\mathcal{P}_{agg}, p_{i_i}^{l-1}, \mathcal{N}_{u_i}^{l-1})
\end{align}

In LightGCN, the initial embeddings of users and items are randomly initialized and updated through backpropagation during training. Similarly, our approach directly utilizes the basic attributes $a_{u_i}$ or $ a_{i_i}$ of users (interacted items and reviews) and items (title, description and  ...) to construct their initial profiles with the pretrained LLM and the system prompt $\mathcal{P}_{init}$ is like

\textit{"You will serve as an assistant to help me determine which types of items a specific user is likely to enjoy.../ summarize which types of users would enjoy a specific item..."}. 

This eliminates the need for additional trainable parameters, making the method more efficient. The generation of initial user/item profile is formulated as follows:
\begin{align}
p_{u_i}^{0} = LLM(\mathcal{P}_{init}, a_{u_i})
p_{i_i}^{0} = LLM(\mathcal{P}_{init}, a_{i_i})
\end{align}

\begin{figure}[t]
\includegraphics[draft=false,width=\textwidth]{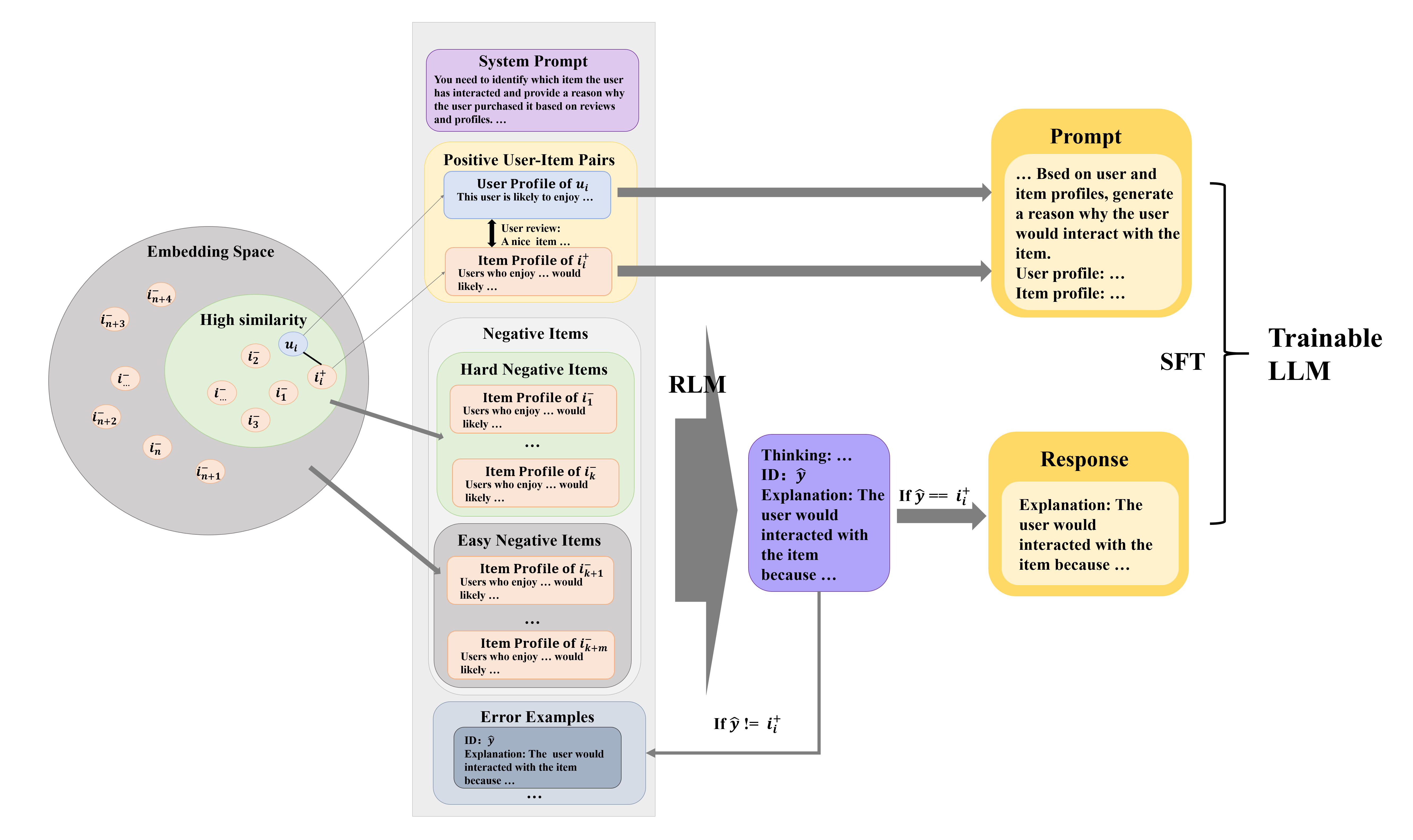}
\caption{An overall overview of the Contrastive Prompting for Explanation Generation and Supervised Fine-Tuning} \label{explanation}
\end{figure}

\subsection{Contrastive Prompting for ground truth Explanation Generation}
To generate high-quality ground truth explanations, we propose a Contrastive Prompting approach. We construct prompts as inputs to a pretrained RLM, enabling it to infer and generate meaningful explanations. The prompt construction is divided into two key components: Discrimination, which helps the model distinguish between relevant and irrelevant items, and Refinement, which enhances explanation quality by leveraging incorrect predictions. An overview of the approach is illustrated in the left part of Figure \ref{explanation}.

\subsubsection{Discrimination.}
We utilize the profiles $p_{u_i}$ and $p_{i_i^+}$ generated by our PGHIS method for the user $u_i$ and the item $i_i^+$ with interaction. Specifically, for each interactive user-item pair, we sample a subset of non-interacted items as negative examples. These negative samples, along with the profiles of the pairs of user-items interacted, are then fed into the pretrained RLM (e.g., DeepSeek-R1\cite{guo2025deepseek}), prompting it to infer which item the user is most likely to engage with and to generate a corresponding explanation. The quality of the explanation is evaluated based on whether the RLM’s predicted item aligns with the actual user-interacted item.

Since incorporating hard negative samples in contrastive learning has been shown to enhance the effectiveness of representation learning\cite{DBLP:conf/iclr/RobinsonCSJ21}, we also introduce hard negatives based on semantic similarity when selecting negative samples. Specifically, given a set $N^-$ of items that have no interaction with user $u_i$, we employ pretrained text encoders to convert both the user profile $p_{u_i}$ and the item profiles in $N^-$ into semantic representations. Using $sim(e_1,e_2) = \frac{e_1 · e_2}{|e_1||e_2|}$ to measure the similarity of two representations $e_1$ and $e_2$, we identify the $k$ items from $N^-$ with the highest semantic similarity to $u_i$ as hard negative samples $i_1^-, .. i_k^-$. Additionally, we randomly sample $m$ items from the remaining items in $N^-$ as standard negative samples $i_{k+1}^- ... i_m^-$. Our initial formula for generating explanations $E$ is as follows:
\begin{equation}
    T, \hat{y}, E = RLM(\mathcal{P}, p_{u_i}, p_{i_i^+}, p_{i_1^-}, ...p_{i_{k+m}^-})
\end{equation}
Here, $T$ refers to the reasoning process of the RLM, while $\hat{y}$ denotes the item that the RLM predicts which the user would interact with. $\mathcal{P}$ refers to the system prompt, which is defined as follows:

\textit{"You will serve as an assistant to help me find the item a specific user has interact with and explain why the user would interact with the item.
I will provide you with the profile of user, review of the item which a specific user has interacted written by the user and a list of items with profiles. ..."}

\subsubsection{Refinement.}
Although existing RLMs possess strong reasoning capabilities, they often fail to produce correct results in a single inference. So we utilize the erroneous reasoning outputs $\hat{Y_{e}}$ and explanations $E_{e}$ generated by the RLM as error examples, incorporating them into the prompt until the model generates the correct item that the user has interacted with. This approach not only helps the model avoid repeating the same mistakes but also provides incorrect explanation references, thereby enhancing the overall quality of the generated explanations. The specific formula is as follows:
\begin{equation}
    T, \hat{y}, E = RLM(\mathcal{P}_{retry}, p_{u_i}, p_{i_i^+}, p_{i_1^-}, ...p_{i_7^-}, \hat{Y_{e}}, E_{e})
\end{equation}
Here, $\mathcal{P}_{retry}$ is derived from $\mathcal{P}$ by incorporating instructions that guide the model to reference incorrect examples.


\subsection{Supervised Fine-Tuning}
\begin{figure}[t]
\centering
\includegraphics[draft=false,width=\textwidth]{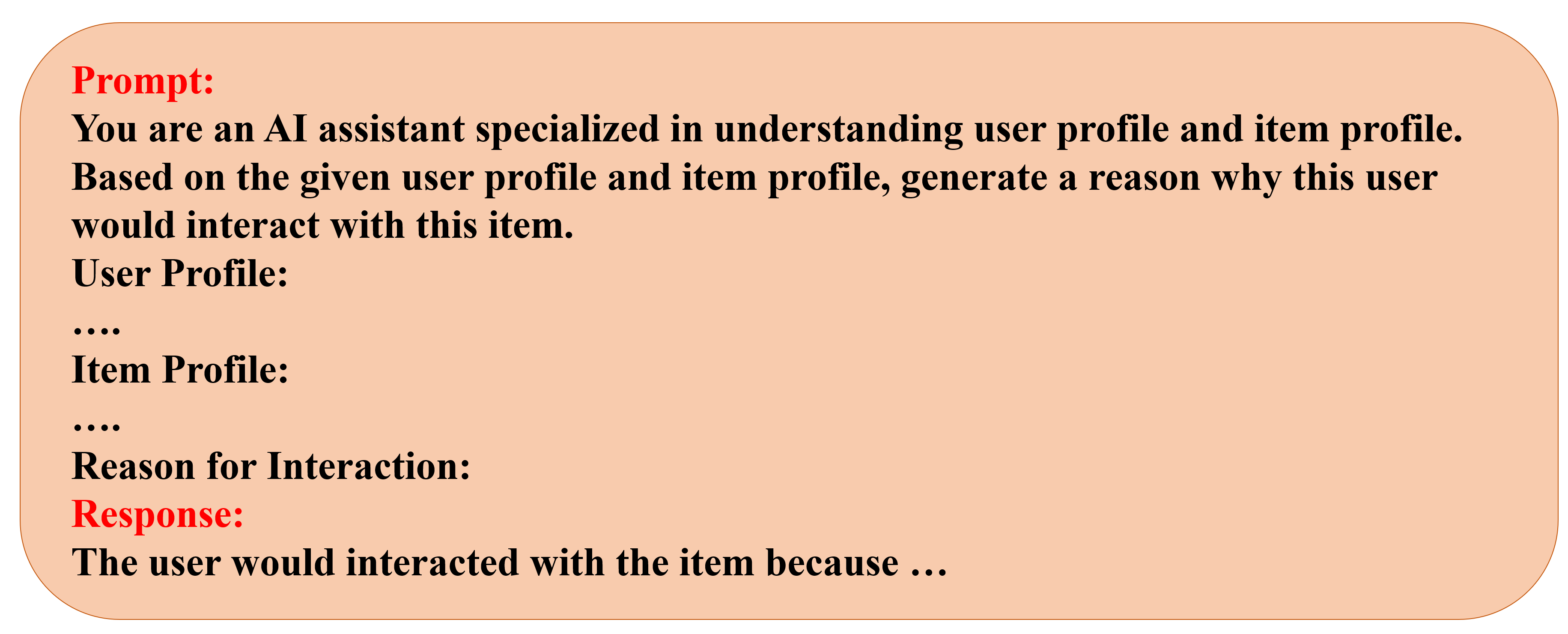}
\caption{Supervised Fine-Tuning dataset format} \label{prompt}
\end{figure}
After obtaining high-quality explanations, we use user and item profiles with interaction as input conditions. Prompt is constructed to instruct the model to generate explanation for user-item interaction based on these profiles, which serves as the input, while the ground truth explanation is used as the output, forming the training samples, as illustrated in Figure \ref{prompt}.

Similar to the standard SFT approach used in decoder-only models, we employ token-level cross-entropy loss as the objective function to update all the parameters of a pretrained LLM. This loss optimizes model parameters by maximizing the probability of predicting the next token given the preceding tokens. The formulation is as follows:
\begin{equation}
    \mathcal{L} = \sum_{t=1}^Tp_{\theta}(x_t|x_1, x_2,...x_{t-1})
\end{equation}
Where $T$ denotes the length of the output sequence, $x_t$ represents the token being predicted, $x_1, x_2, ..., x_{t-1}$ refers to the preceding tokens in the sequence, $\theta$ denotes the trainable model parameters, and $p_{\theta}(x_t|x_1, x_2,...x_{t-1})$ represents the predicted probability distribution of the model.

\section{Experiments}
\subsection{Experimental Seting}
\subsubsection{Datasets.} 
We evaluated our approach using three widely used public datasets. \textbf{Amazon-Book} \footnote{https://nijianmo.github.io/amazon/index.html}, comprises user ratings and reviews for books available on Amazon. \textbf{Yelp} \footnote{https://www.yelp.com/dataset}, captures user interactions with businesses and provides detailed category information across various industries. \textbf{Steam}\cite{DBLP:conf/icdm/KangM18} contains user feedback and engagement data related to video games on the Steam platform. As the Steam dataset does not contain rating scores, no filtering was applied. For the other datasets, we first applied k-core filtering and then divided all of them into three subsets using a 3:1:1 ratio. These subsets were used for profile generation, sampling 20,000 instances for SFT training, and selecting 2,000 instances for testing.

\subsubsection{Evaluation Metrics.}
Following the evaluation approach of XRec\cite{DBLP:conf/emnlp/MaR024}, we have chosen several advanced metrics to rigorously assess the explainability of the explanations generated by our trained model. \textbf{GPTScore}\cite{DBLP:conf/naacl/FuNJ024} serves as a scalable alternative to human evaluation, demonstrating strong correlations with expert judgments in text generation task. \textbf{BERTScore}\cite{DBLP:conf/iclr/ZhangKWWA20} is a semantic evaluation metric that measures the similarity between candidate and reference texts using contextualized embeddings from pretrained BERT models. \textbf{BLEURT}\cite{DBLP:conf/acl/SellamDP20} is a learned evaluation metric that leverages BERT-based contextual embeddings and fine-tuning on human-rated data to assess text quality.

Although traditional text generation evaluation metrics struggle to effectively capture semantic similarity between generated text and ground truth, we nonetheless employ certain metrics as a reference for assessing the text quality. \textbf{BLEU} (Bilingual Evaluation Understudy)\cite{DBLP:conf/acl/PapineniRWZ02} measures n-gram overlap between generated and reference texts, applying brevity penalties to account for length differences. \textbf{ROUGE} (Recall-Oriented Understudy for Gisting Evaluation)\cite{lin-2004-rouge} evaluates text summarization by measuring lexical overlap between generated and reference summaries.

\subsubsection{Compared Method.}
\begin{table}[t] 
	\caption{Overall performance comparison with baselines on Amazon-Book, Yelp and Steam datasets. ↑ indicates that higher values correspond to better performance. "BS" denotes BERTScore, while "P," "R," and "F1" represent Precision, Recall, and F1-Score, respectively. "R-1," "R-2," and "R-L" refer to ROUGE-1, ROUGE-2, and ROUGE-L scores. The best-performing results are highlighted in bold, while the second-best results are underlined for emphasis.}
	\centering
	\label{res}
        \setlength{\tabcolsep}{3.5pt}
        \resizebox{1.0\textwidth}{!}{
        \begin{tabular}{ccccccccccc}
		\toprule
		\multirow{2}{*}{\centering Methods} & \multicolumn{5}{c}{Explainability↑} & \multicolumn{5}{c}{Text Quality↑} \\
		\cmidrule(lr){2-6}\cmidrule(lr){7-11}
            &  GPTScore &  BS\textsuperscript{P} &  BS\textsuperscript{R} &  BS\textsuperscript{F1} &  BLEURT &  BLEU-1 &  BLEU-4 &  R-1 &  R-2 &   R-L \\
		\midrule
            \multicolumn{11}{c}{Amazon-Books} \\
            \midrule
            Att2Seq&  71.75 & 0.6119 &  0.6026 &  0.6076 &  0.6086 &  0.5959 &  0.3535 &  0.5736 &  0.3835 &  0.5110 \\
            NRT &  \underline{84.58} & 0.2814 &  0.6449 &  0.4577 &  0.6088 &  0.2472 &  0.1047 &  0.3755 &  0.2087 &  0.2860 \\
            PETER&  69.10 & 0.6145 &  0.6198 &  0.6175 &  0.6143 &  0.5884 &  0.3550 &  0.5833 &  0.3922 &  0.5216 \\
            PEPLER&  77.54 & 0.6173 &  0.6152 &  0.6165 &  0.5900 &  \underline{0.6097} &  \underline{0.3705} &  \underline{0.5996} &  \underline{0.4059} &  \underline{0.5291} \\
            XRec&  83.76 & \underline{0.6528} &  \underline{0.6751} &  \underline{0.6642} &  \underline{0.6300} &  0.6018 &  0.3386 &  0.5976 &  0.3927 &  0.5191 \\
            Ours&  \textbf{88.24} & \textbf{0.7209} &  \textbf{0.7294} &  \textbf{0.7253} &  \textbf{0.6732} &  \textbf{0.6947} &  \textbf{0.4555} &  \textbf{0.6816} &  \textbf{0.5045} &  \textbf{0.6105} \\
            \midrule

            \multicolumn{11}{c}{Yelp} \\
            \midrule
            Att2Seq&  54.60 & 0.5459 &  0.5311 &  0.5389 & 0.5153  &  0.5704 &  0.2769 &  0.5411 &  0.3028 &  0.4629 \\
            NRT &  \underline{83.99} & 0.3035 &  0.6441 &  0.4691 &  0.4467 &  0.2648 &  0.1038 &  0.3889 &  0.2009 &  0.2878 \\
            PETER&  69.13 & 0.6034 &  0.5859 &  0.5949 & 0.5697  &  \underline{0.6284} &  \underline{0.3418} &  0.6018 &  0.3654 &  \underline{0.5156} \\
            PEPLER&  72.78 & 0.6047 &  0.5813 &  0.5932 & 0.5376  &  0.5987 &  0.3364 &  0.5979 &  \underline{0.3711} &  0.5146 \\
            XRec&  83.30 & \underline{0.6272} &  \underline{0.6658} &  \underline{0.6466} &  \underline{0.5905} &  0.5922 &  0.3045 &  \underline{0.6065} &  0.3635 &  0.5132 \\
            Ours&  \textbf{88.83} & \textbf{0.7277} &  \textbf{0.7394} &  \textbf{0.7336} &  \textbf{0.6639} &  \textbf{0.7298} &  \textbf{0.4864} &  \textbf{0.7102} &  \textbf{0.5208} &  \textbf{0.6354} \\
            \midrule

            \multicolumn{11}{c}{Steam} \\
            \midrule
            Att2Seq&  58.56& 0.6269 &  0.6114 &  0.6195 & 0.6251  &  0.6304 &  0.4004 &  0.6100 &  0.4190 &  0.5534 \\
            NRT &  61.29 & 0.6379 &  0.6077 &  0.6230 &  0.6213 &  0.6212 &  0.3992 &  0.6119 &  0.4248 &  0.5602 \\
            PETER&  69.15 & \underline{0.6497} &  0.6380 &  0.6441 & \underline{0.6342}  &  0.6527 &  \underline{0.4218} &   \underline{0.6375} &  \underline{0.4437} &  \underline{0.5758} \\
            PEPLER&  67.67 & 0.6491 &  \underline{0.6396} &  \underline{0.6447} &  0.6171 &  \underline{0.6571} &  0.4208 & 0.6354 &  0.4412 &  0.5741 \\
            XRec&  \underline{74.38} & 0.5593 &  0.5803 &  0.5702 &  0.5135 &  0.5450 &  0.2142 &  0.5392 &  0.2946 &  0.4611 \\
            Ours&  \textbf{81.55} & \textbf{0.7315} &  \textbf{0.7269} &  \textbf{0.7294} &  \textbf{0.6728} &  \textbf{0.7301} &  \textbf{0.4946} &  \textbf{0.7104} &  \textbf{0.5276} &  \textbf{0.6448} \\
		
		\bottomrule
	\end{tabular}
        }
\end{table}
We compared our method with the following five state-of-the-art baselines:
\begin{itemize}
    \item \textbf{Att2Seq}\cite{DBLP:conf/eacl/ZhouLWDHX17} proposes an attention-enhanced attribute-to-sequence model that generates personalized item reviews based on user, item, and rating attributes. For steam dataset, we use playing hours instead of rating.
    \item \textbf{NRT}\cite{DBLP:conf/sigir/LiWRBL17} proposes a neural framework that jointly predicts user ratings and generates concise textual tips by integrating collaborative filtering with a gated recurrent neural network decoder.
    \item \textbf{PETER}\cite{DBLP:conf/acl/LiZC20} proposes a transformer-based model that generates personalized, explainable recommendations by integrating user and item IDs into text generation. 
    \item \textbf{PEPLER}\cite{DBLP:journals/tois/LiZC23} proposes a personalized prompt learning approach for explainable recommendation, integrating user and item identifiers into pretrained language models to generate tailored explanations. 
    \item \textbf{XRec}\cite{DBLP:conf/emnlp/MaR024} employs a lightweight collaborative adaptor to incorporate collaborative signals, enabling large language models to understand complex user-item interactions and provide comprehensive explanations for user behaviors in recommender systems.
\end{itemize}
\subsubsection{Implementation Details.}
Due to constraints on prompt length and computational cost, we select neighboring nodes based on rating or playing hours, prioritizing the top 15 for profile generation using GPT-4o-mini\footnote{https://openai.com/index/hello-gpt-4o/} and use the profiles generated by RLMRec\cite{DBLP:conf/www/RenWXSCWY024} as initial profiles. For explanation generation, we use all-MiniLM-L6-v2\cite{minilmv2} to generate embeddings of profiles and  employ DeepSeek-R1\cite{guo2025deepseek} as the RLM and set $k$ and $m$ to 2 and 5, respectively. Finally, we fine-tune a cost-effective Qwen2.5-7B-Instruction\cite{qwen2.5} model with a learning rate of 1e-5 on a single H20 GPU, using LLaMA-Factory\cite{zheng2024llamafactory} to full-parameter SFT our model.

\subsection{Performance Comparison}
The results across the Amazon-Books, Yelp, and Steam datasets showed in table \ref{res} demonstrates the superior performance of our method compared to established baselines in both explainability and text quality metrics. 
\begin{itemize}
    \item For explainability, our method exhibits significant enhancements across five explainability metrics. Notably, when evaluated using the GPTScore metric derived from the state-of-the-art GPT-3.5 model, our method surpasses the second-best approach by an average of 5 points out of 100 across three datasets. This performance underscores our method's ability to thoroughly leverage user and item characteristics, thereby generating high-quality explanations.
    \item For text quality, our method achieves superior text quality across all datasets, outperforming baselines in ROUGE-1, ROUGE-2, and ROUGE-L. For example, on Amazon-Books, our method surpasses the second-best approach, PEPLER, by nearly 0.1 across all three ROUGE metrics. Strong BLEU scores further validate its fluency and accuracy, demonstrating the effectiveness of our approach in generating high-quality explanations.
\end{itemize}
The superior performance of our method can be attributed to two key factors. (1) Unlike approaches that rely on learned user or item embeddings, our method explicitly represents user preferences and interactions in textual form. This textual information serves as input to LLM , making it more interpretable and effective for learning compared to embeddings. (2) We fine-tune a general LLM through SFT, enabling it to generate more coherent and high-quality explanations.
\begin{figure}[t]
\centering
\includegraphics[draft=false,width=\textwidth]{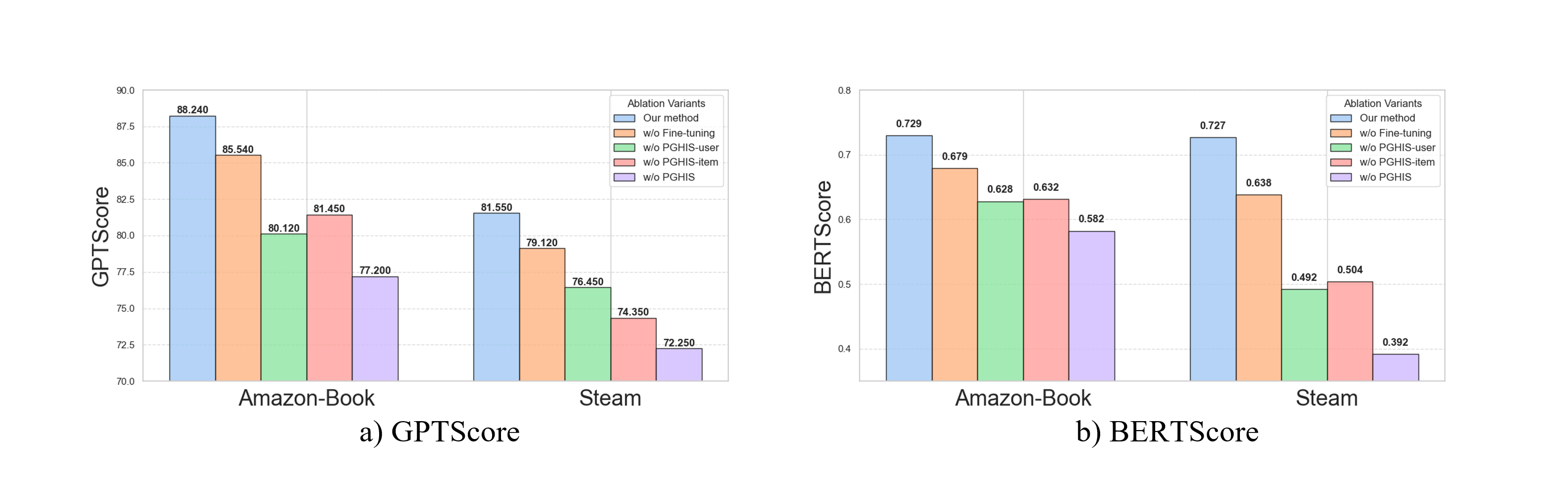}
\caption{Ablastion Study. a) and b) respectively illustrate the ablation study results of GPTScore and BERTScore on two datasets.} \label{ablation}
\end{figure}
\subsection{Ablation Study}
To evaluate the contribution of each component in our approach, we conducted an ablation study on GPTScore and BERTScore by comparing \textbf{i) Our method} with four variants: \textbf{ii) w/o Fine-tuning}, which generates explanations using our profiles without SFT; \textbf{iii) w/o PGHIS-user}, which replaces PGHIS-generated user profiles with initial ones while retaining PGHIS-generated item profiles; \textbf{iv) w/o PGHIS-item}, which removes PGHIS-generated item profiles; and \textbf{v) w/o PGHIS}, which uses only the initial user and item profiles. The results are shown in Figure \ref{ablation}.

The results indicate that without the use of SFT, both GPTScore and BERTScore only experience slight declines, with performance still surpassing that of XRec. This suggests that the profiles generated by PGHIS effectively capture the preferences and interaction information of users and items, enabling the model to generate high quality explanations even without SFT. Furthermore, when considering the results from using either the item or user profiles generated solely by PGHIS, it is evident that both have a nearly equivalent level of importance in the outcome. In contrast, when PGHIS-generated profiles are not utilized, both GPTScore and BERTScore experience a significant decrease, highlighting the superiority of PGHIS.

\subsection{Ground Truth Explanation Quality}
\begin{figure}[t]
\centering
\includegraphics[draft=false,width=\textwidth]{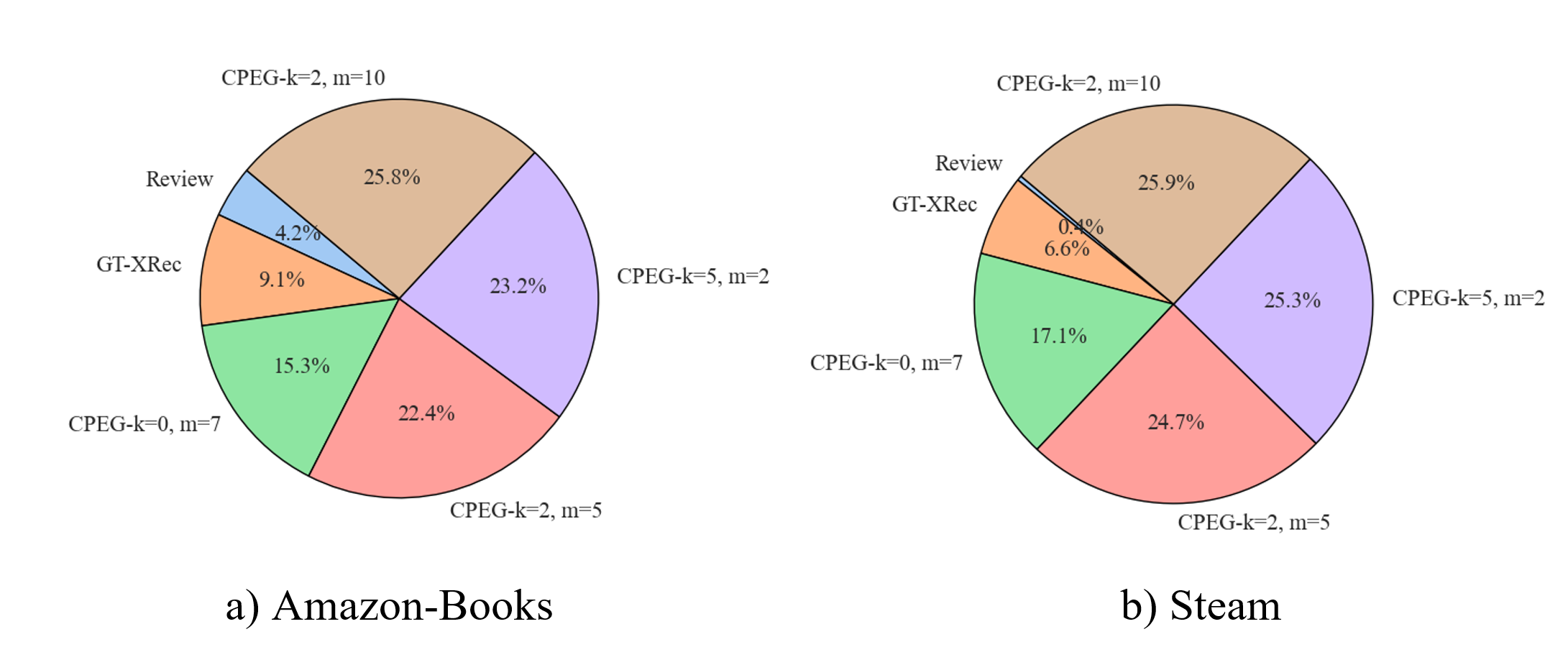}
\caption{Win Rate Comparison of Ground Truths in Explanation Quality. a) illustrates the win rates of different ground truth explanations in the Amazon-Books dataset, while b) presents the corresponding results for the Steam dataset.} \label{gt}
\end{figure}
Given that the quality assessment of ground truth is inherently subjective. we refer to existing studies on utilizing LLMs-as-a-judge\cite{DBLP:journals/corr/abs-2411-15594} to evaluate the quality of ground truth explanations for user-item interactions and assess the impact of $k$ and $m$ on explanation quality, we conduct experiments on the Amazon-Books and Steam datasets. Using user-item profiles with interaction as a reference, GPT-3.5 determines the most appropriate explanation among: \textbf{i) Review}: user-written review, \textbf{ii) GT-XRec}: ground truth generated by XRec, \textbf{iii) CPEG-k=0, m=7}: CPEG without hard negative samples and m=7, \textbf{iv) CPEG-k=2, m=5}: CPEG with k=2 and m=5, \textbf{v) CPEG-k=5, m=2}: CPEG with k=5 and m=2, and \textbf{vi) CPEG-k=2, m=10}: CPEG with k=2 and m=10. The win rates of each ground truth explanation are shown in Figure \ref{gt}.

The results indicate that ground truths generated by CPEG significantly outperform both user-written reviews and those produced by XRec\cite{DBLP:conf/emnlp/MaR024}, achieving win rates of 86.7\% and 93.0\% on the two datasets, respectively. Compared to using only 7 randomly selected negative samples, incorporating 2 hard negatives and 5 random negatives improves win rates by 7.1\% and 7.6\%, highlighting the importance of hard negatives in generating high-quality explanations. Similar to contrastive learning in representation learning\cite{DBLP:conf/icml/ChenK0H20}, increasing the number of hard and random negatives enhances explanation quality. However, for RLM, this makes identifying the correct answer increasingly difficult, leading to more retries and higher computational costs while yielding diminishing improvements. Based on this trade-off, we set $k = 2$ and $m = 5$ in our experiments.

\subsection{Case Study}
\begin{figure}[t]
\centering
\includegraphics[draft=false,width=\textwidth]{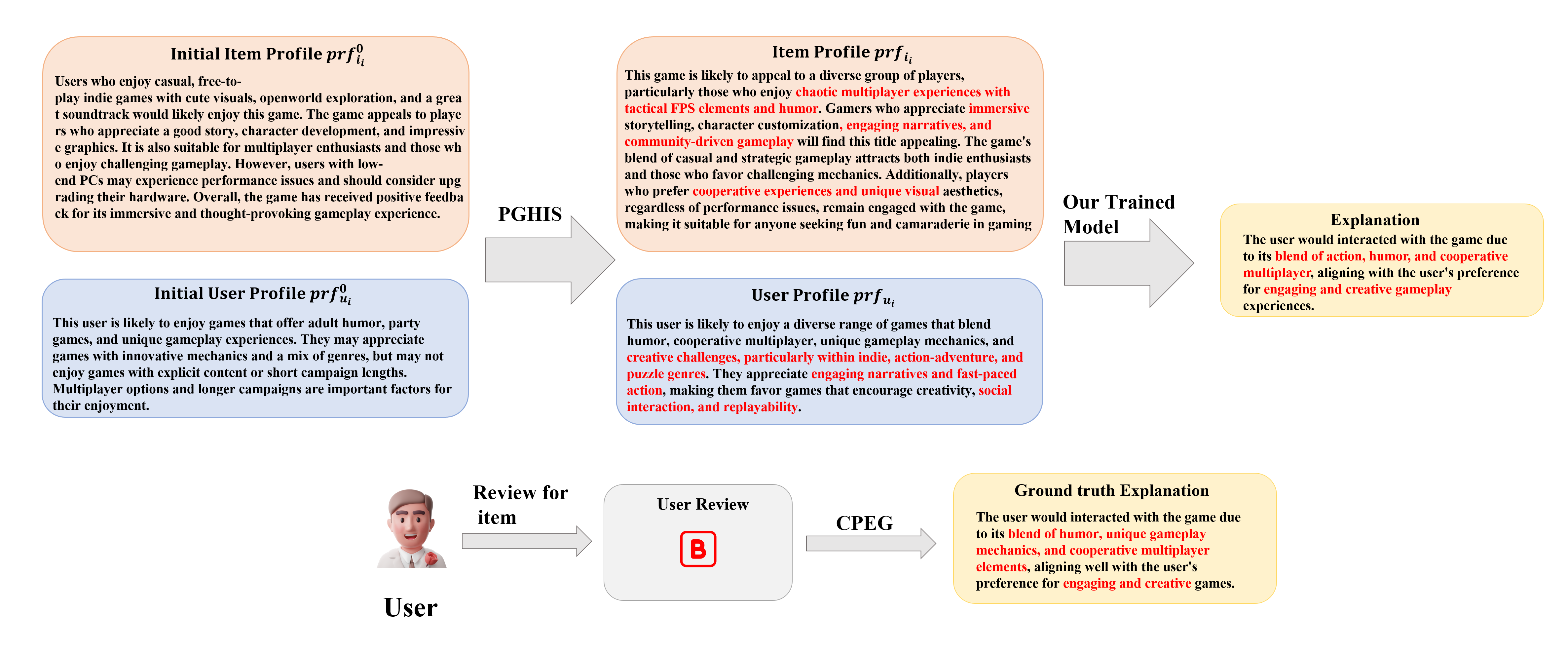}
\caption{Case Study} \label{case}
\end{figure}

To intuitively demonstrate our approach, Figure \ref{case} presents a case example from a test set. As illustrated in the example, on the one hand, the initial user and item profiles, after being processed by PGHIS, contain significantly more preference information than the original profiles. For instance, the item profile now includes details such as "chaotic multiplayer experiences with tactical FPS elements and humor," which were not part of the initial profile. On the other hand, despite the user's comment consisting only of an emoji without any explicit reasons for their preference, CPEG is still able to generate a reasonable explanation, attributing the interaction between the user and the item. After SFT, when given user and item profiles, our model generates explanations that are nearly identical to the ground truth, highlighting the superiority of our approach.
\section{Conclusion}
In this work, we propose Profile Generation via Hierarchical Interaction Summarization, which replaces traditional embeddings with explicit textual representations of user and item characteristics. Additionally, we introduce Contrastive Prompting for ground truth Explanation Generation, a method designed to enhance the quality of generated explanations by guiding LLMs with both positive and negative items. Our approach was evaluated against state-of-the-art baselines across diverse datasets, demonstrating an average improvement of 5/100 on GPTScore, 20.6\% and 19.6\% on variants of BLEU and ROUGE over the state-of-the-art methods. The experimental results of ground truth explanation quality comparison further confirm the superior quality of our generated ground truth explanations. 

In future work, we aim to apply our profile aggregation approach to LLM-based recommender systems, improving recommendation explainability and performence. Additionally, we hope that CPEG can assist researchers in generating high-quality data for studies requiring rigorous reasoning processes and explanations.

\section{Acknowledgments}
This research was supported by STI 2030—Major Projects under Grant 2021ZD0200403 and Zhejiang Provincial Natural Science Foundation of China under Grant No.LD24F030002. The authors like to thank the anonymous reviewers for their review and comments.

\begin{thebibliography}{10}
\providecommand{\url}[1]{\texttt{#1}}
\providecommand{\urlprefix}{URL }
\providecommand{\doi}[1]{https://doi.org/#1}

\bibitem{besta2025reasoning}
Besta, M., Barth, J., Schreiber, E., Kubicek, A., Catarino, A., Gerstenberger, R., Nyczyk, P., Iff, P., Li, Y., Houliston, S., et~al.: Reasoning language models: A blueprint. arXiv preprint arXiv:2501.11223  (2025)

\bibitem{DBLP:journals/corr/abs-2412-18925}
Chen, J., Cai, Z., Ji, K., Wang, X., Liu, W., Wang, R., Hou, J., Wang, B.: Huatuogpt-o1, towards medical complex reasoning with llms. CoRR  \textbf{abs/2412.18925} (2024)

\bibitem{DBLP:conf/icml/ChenK0H20}
Chen, T., Kornblith, S., Norouzi, M., Hinton, G.E.: A simple framework for contrastive learning of visual representations. In: {ICML}. Proceedings of Machine Learning Research, vol.~119, pp. 1597--1607. {PMLR} (2020)

\bibitem{DBLP:journals/corr/abs-2311-09277}
Chia, Y.K., Chen, G., Tuan, L.A., Poria, S., Bing, L.: Contrastive chain-of-thought prompting. CoRR  \textbf{abs/2311.09277} (2023)

\bibitem{DBLP:journals/corr/ChungGCB14}
Chung, J., G{\"{u}}l{\c{c}}ehre, {\c{C}}., Cho, K., Bengio, Y.: Empirical evaluation of gated recurrent neural networks on sequence modeling. CoRR  \textbf{abs/1412.3555} (2014), \url{http://arxiv.org/abs/1412.3555}

\bibitem{DBLP:journals/corr/abs-2109-09587}
Deng, Y.: Recommender systems based on graph embedding techniques: {A} comprehensive review. CoRR  \textbf{abs/2109.09587} (2021), \url{https://arxiv.org/abs/2109.09587}

\bibitem{DBLP:conf/eacl/ZhouLWDHX17}
Dong, L., Huang, S., Wei, F., Lapata, M., Zhou, M., Xu, K.: Learning to generate product reviews from attributes. In: {EACL} {(1)}. pp. 623--632. Association for Computational Linguistics (2017)

\bibitem{DBLP:conf/naacl/FuNJ024}
Fu, J., Ng, S., Jiang, Z., Liu, P.: Gptscore: Evaluate as you desire. In: {NAACL-HLT}. pp. 6556--6576. Association for Computational Linguistics (2024)

\bibitem{DBLP:journals/corr/abs-2411-15594}
Gu, J., Jiang, X., Shi, Z., Tan, H., Zhai, X., Xu, C., Li, W., Shen, Y., Ma, S., Liu, H., Wang, Y., Guo, J.: A survey on llm-as-a-judge. CoRR  \textbf{abs/2411.15594} (2024)

\bibitem{guo2025deepseek}
Guo, D., Yang, D., Zhang, H., Song, J., Zhang, R., Xu, R., Zhu, Q., Ma, S., Wang, P., Bi, X., et~al.: Deepseek-r1: Incentivizing reasoning capability in llms via reinforcement learning. arXiv preprint arXiv:2501.12948  (2025)

\bibitem{DBLP:conf/sigir/0001DWLZ020}
He, X., Deng, K., Wang, X., Li, Y., Zhang, Y., Wang, M.: Lightgcn: Simplifying and powering graph convolution network for recommendation. In: Huang, J.X., Chang, Y., Cheng, X., Kamps, J., Murdock, V., Wen, J., Liu, Y. (eds.) Proceedings of the 43rd International {ACM} {SIGIR} conference on research and development in Information Retrieval, {SIGIR} 2020, Virtual Event, China, July 25-30, 2020. pp. 639--648. {ACM} (2020). \doi{10.1145/3397271.3401063}, \url{https://doi.org/10.1145/3397271.3401063}

\bibitem{DBLP:conf/kdd/HouMZLDW22}
Hou, Y., Mu, S., Zhao, W.X., Li, Y., Ding, B., Wen, J.: Towards universal sequence representation learning for recommender systems. In: Zhang, A., Rangwala, H. (eds.) {KDD} '22: The 28th {ACM} {SIGKDD} Conference on Knowledge Discovery and Data Mining, Washington, DC, USA, August 14 - 18, 2022. pp. 585--593. {ACM} (2022). \doi{10.1145/3534678.3539381}, \url{https://doi.org/10.1145/3534678.3539381}

\bibitem{DBLP:conf/icdm/KangM18}
Kang, W.C., McAuley, J.: Self-attentive sequential recommendation. In: 2018 IEEE international conference on data mining (ICDM). pp. 197--206. IEEE (2018)

\bibitem{DBLP:conf/acl/LiZC20}
Li, L., Zhang, Y., Chen, L.: Personalized transformer for explainable recommendation. In: {ACL/IJCNLP} {(1)}. pp. 4947--4957. Association for Computational Linguistics (2021)

\bibitem{DBLP:journals/tois/LiZC23}
Li, L., Zhang, Y., Chen, L.: Personalized prompt learning for explainable recommendation. {ACM} Trans. Inf. Syst.  \textbf{41}(4),  103:1--103:26 (2023)

\bibitem{li2024learning}
Li, M., Aggarwal, K., Xie, Y., Ahmad, A., Lau, S.: Learning from contrastive prompts: Automated optimization and adaptation. arXiv preprint arXiv:2409.15199  (2024)

\bibitem{DBLP:conf/sigir/LiWRBL17}
Li, P., Wang, Z., Ren, Z., Bing, L., Lam, W.: Neural rating regression with abstractive tips generation for recommendation. In: {SIGIR}. pp. 345--354. {ACM} (2017)

\bibitem{lin-2004-rouge}
Lin, C.Y.: {ROUGE}: A package for automatic evaluation of summaries. In: Text Summarization Branches Out. pp. 74--81. Association for Computational Linguistics, Barcelona, Spain (Jul 2004), \url{https://aclanthology.org/W04-1013/}

\bibitem{DBLP:conf/emnlp/MaR024}
Ma, Q., Ren, X., Huang, C.: Xrec: Large language models for explainable recommendation. In: {EMNLP} (Findings). pp. 391--402. Association for Computational Linguistics (2024)

\bibitem{DBLP:conf/acl/PapineniRWZ02}
Papineni, K., Roukos, S., Ward, T., Zhu, W.: Bleu: a method for automatic evaluation of machine translation. In: {ACL}. pp. 311--318. {ACL} (2002)

\bibitem{DBLP:conf/www/RenWXSCWY024}
Ren, X., Wei, W., Xia, L., Su, L., Cheng, S., Wang, J., Yin, D., Huang, C.: Representation learning with large language models for recommendation. In: {WWW}. pp. 3464--3475. {ACM} (2024)

\bibitem{DBLP:conf/iclr/RobinsonCSJ21}
Robinson, J.D., Chuang, C., Sra, S., Jegelka, S.: Contrastive learning with hard negative samples. In: {ICLR}. OpenReview.net (2021)

\bibitem{DBLP:journals/jbd/RoyD22}
Roy, D., Dutta, M.: A systematic review and research perspective on recommender systems. J. Big Data  \textbf{9}(1), ~59 (2022). \doi{10.1186/S40537-022-00592-5}, \url{https://doi.org/10.1186/s40537-022-00592-5}

\bibitem{DBLP:conf/acl/SellamDP20}
Sellam, T., Das, D., Parikh, A.P.: {BLEURT:} learning robust metrics for text generation. In: {ACL}. pp. 7881--7892. Association for Computational Linguistics (2020)

\bibitem{DBLP:conf/nips/ShiCWYWW15}
Shi, X., Chen, Z., Wang, H., Yeung, D., Wong, W., Woo, W.: Convolutional {LSTM} network: {A} machine learning approach for precipitation nowcasting. In: Cortes, C., Lawrence, N.D., Lee, D.D., Sugiyama, M., Garnett, R. (eds.) Advances in Neural Information Processing Systems 28: Annual Conference on Neural Information Processing Systems 2015, December 7-12, 2015, Montreal, Quebec, Canada. pp. 802--810 (2015), \url{https://proceedings.neurips.cc/paper/2015/hash/07563a3fe3bbe7e3ba84431ad9d055af-Abstract.html}

\bibitem{DBLP:conf/nips/VaswaniSPUJGKP17}
Vaswani, A., Shazeer, N., Parmar, N., Uszkoreit, J., Jones, L., Gomez, A.N., Kaiser, L., Polosukhin, I.: Attention is all you need. In: Guyon, I., von Luxburg, U., Bengio, S., Wallach, H.M., Fergus, R., Vishwanathan, S.V.N., Garnett, R. (eds.) Advances in Neural Information Processing Systems 30: Annual Conference on Neural Information Processing Systems 2017, December 4-9, 2017, Long Beach, CA, {USA}. pp. 5998--6008 (2017), \url{https://proceedings.neurips.cc/paper/2017/hash/3f5ee243547dee91fbd053c1c4a845aa-Abstract.html}

\bibitem{minilmv2}
Wang, W., Bao, H., Huang, S., Dong, L., Wei, F.: {M}ini{LM}v2: Multi-head self-attention relation distillation for compressing pretrained transformers. In: Findings of the Association for Computational Linguistics: ACL-IJCNLP 2021. pp. 2140--2151. Association for Computational Linguistics, Online (Aug 2021). \doi{10.18653/v1/2021.findings-acl.188}, \url{https://aclanthology.org/2021.findings-acl.188}

\bibitem{NGCF19}
Wang, X., He, X., Wang, M., Feng, F., Chua, T.: Neural graph collaborative filtering. In: Proceedings of the 42nd International {ACM} {SIGIR} Conference on Research and Development in Information Retrieval, {SIGIR} 2019, Paris, France, July 21-25, 2019. pp. 165--174 (2019)

\bibitem{autocf2023}
Xia, L., Huang, C., Huang, C., Lin, K., Yu, T., Kao, B.: Automated self-supervised learning for recommendation. In: The Web Conference (WWW) (2023)

\bibitem{qwen2.5}
Yang, A., Yang, B., Zhang, B., Hui, B., Zheng, B., Yu, B., Li, C., Liu, D., Huang, F., Wei, H., Lin, H., Yang, J., Tu, J., Zhang, J., Yang, J., Yang, J., Zhou, J., Lin, J., Dang, K., Lu, K., Bao, K., Yang, K., Yu, L., Li, M., Xue, M., Zhang, P., Zhu, Q., Men, R., Lin, R., Li, T., Xia, T., Ren, X., Ren, X., Fan, Y., Su, Y., Zhang, Y., Wan, Y., Liu, Y., Cui, Z., Zhang, Z., Qiu, Z.: Qwen2.5 technical report. arXiv preprint arXiv:2412.15115  (2024)

\bibitem{DBLP:conf/iclr/ZhangKWWA20}
Zhang, T., Kishore, V., Wu, F., Weinberger, K.Q., Artzi, Y.: Bertscore: Evaluating text generation with {BERT}. In: {ICLR}. OpenReview.net (2020)

\bibitem{zheng2024llamafactory}
Zheng, Y., Zhang, R., Zhang, J., Ye, Y., Luo, Z., Feng, Z., Ma, Y.: Llamafactory: Unified efficient fine-tuning of 100+ language models. In: Proceedings of the 62nd Annual Meeting of the Association for Computational Linguistics (Volume 3: System Demonstrations). Association for Computational Linguistics, Bangkok, Thailand (2024), \url{http://arxiv.org/abs/2403.13372}

\end{thebibliography}
\end{document}